\documentclass{elsart}
\usepackage{graphicx} % Include figure files
\usepackage{color}
\usepackage{cite}
\usepackage{epsf}

\newcommand{\cal}{\mathcal }
\newcommand{\notM}{\mu\!\!\!/}
\newcommand{\notE}{e\!\!\!/}
\newcommand{\TMG}{\tau\rightarrow\mu\gamma}
\newcommand{\TEG}{\tau\rightarrow{e}\gamma}
\newcommand{\MG}{\mu\gamma}
\newcommand{\EG}{{e}\gamma}
\newcommand{\fbi}{fb${}^{-1}$}

\newcommand{\gev}{{GeV}}

\newcommand{\gevpcs}{{GeV$/c^2$}}
\newcommand{\mev}{{MeV}}

\newcommand{\mevpcs}{{MeV$/c^2$}}

\newcommand{\notmu}{\mu\!\!\!/}
\newcommand{\minv}{M_{\rm inv}}
\newcommand{\dE}{{\it \Delta} E}

\begin{document}
\begin{frontmatter}
\epsfysize3cm
\epsfbox{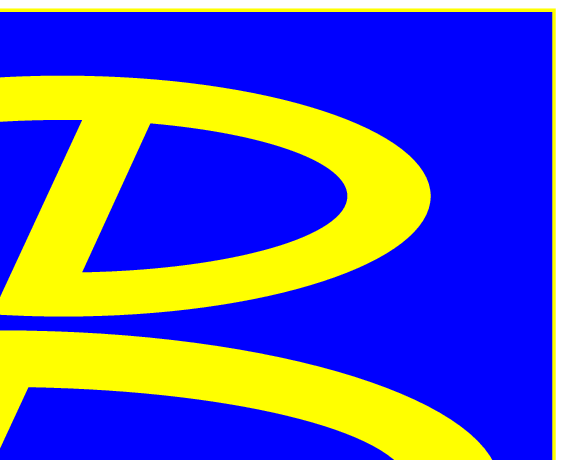}\hspace*{\fill}    % BELLE-logo (you need file belle.eps)
\vskip -3cm
\noindent
\hspace*{3.7in}Belle Preprint 2007-6 \\
\hspace*{3.7in}KEK Preprint 2006-69 \\
\vskip 1.5cm
\title{ 
New Search for $\tau \to \mu \gamma $ and $\tau \to e \gamma$ Decays at Belle
}
\vskip -1.0cm
%%% Paper:    tau -> l gamma
%%% Journal:  Physics Letters B
%%% Contacts: K. Hayasaka (hayasaka@hepl.phys.nagoya-u.ac.jp)
%%% Non-responding authors or those who said NO are commented out.
%%% ====================================================================
%%% Click the RELOAD button on your web browser to see the updated file.
%%% ====================================================================
%%% Use \input{author} to insert this material into your latex file.
\collab{Belle Collaboration}
  \author[Nagoya]{K.~Hayasaka}, % Nagoya
% \author[KEK]{K.~Abe}, % KEK
  \author[TohokuGakuin]{K.~Abe}, % TohokuGakuin
% \author[TIT]{N.~Abe}, % TIT
  \author[KEK]{I.~Adachi}, % KEK
  \author[Tokyo]{H.~Aihara}, % Tokyo
  \author[BINP]{D.~Anipko}, % BINP
% \author[Nagoya]{K.~Aoki}, % Nagoya
  \author[BINP]{K.~Arinstein}, % BINP
% \author[Tsukuba]{Y.~Asano}, % Tsukuba
% \author[Toyama]{T.~Aso}, % Toyama
  \author[BINP]{V.~Aulchenko}, % BINP
% \author[Lausanne,ITEP]{T.~Aushev}, % ITEP
% \author[Tata]{T.~Aziz}, % Tata
% \author[Cincinnati]{S.~Bahinipati}, % Cincinnati
  \author[Sydney]{A.~M.~Bakich}, % Sydney
% \author[ITEP]{V.~Balagura}, % ITEP
% \author[Peking]{Y.~Ban}, % Peking
% \author[Tata]{S.~Banerjee}, % Tata
  \author[Melbourne]{E.~Barberio}, % Melbourne
% \author[Hawaii]{M.~Barbero}, % Hawaii
  \author[Lausanne]{A.~Bay}, % Lausanne
% ...
% Ljubljana
% \author[Lausanne]{D.~Z\"urcher}, % Lausanne

\address[BINP]{Budker Institute of Nuclear Physics, Novosibirsk, Russia}
% ...
\address[KEK]{High Energy Accelerator Research Organization (KEK), Tsukuba, Japan}
% ...
\address[Lausanne]{Swiss Federal Institute of Technology of Lausanne, EPFL, Lausanne, Switzerland}
% ...
\address[Melbourne]{University of Melbourne, Victoria, Australia}
\address[Nagoya]{Nagoya University, Nagoya, Japan}
% ...
\address[Sydney]{University of Sydney, Sydney, NSW, Australia}
%%%\address[Tata]{Tata Institute of Fundamental Research, Bombay, India}
% ...
\address[TohokuGakuin]{Tohoku Gakuin University, Tagajo, Japan}
% ...
\address[Tokyo]{Department of Physics, University of Tokyo, Tokyo, Japan}
% ...

\begin{abstract}
We report on a search for the lepton flavor violating 
$\tau^- \to \mu^- \gamma$ 
and $\tau^- \to e^- \gamma$ decays based on 535 fb$^{-1}$ 
of data accumulated at the Belle experiment. 
No signal is found and
we set 90\% confidence level upper limits on
the branching ratios
${\cal B}(\tau^- \to \mu^- \gamma) < 4.5\times 10^{-8}$ and
${\cal B}(\tau^- \to e^- \gamma) < 1.2\times 10^{-7}$.
\end{abstract}

\vspace*{-7mm}
\begin{keyword}
Decays of taus\sep lepton number\sep Taus
\PACS{13.35.Dx, 11.30.Fs, 14.60.Fg}
\end{keyword}
\end{frontmatter}

\section{Introduction}
Lepton flavor violation (LFV) could appear in various new physics
scenarios beyond the Standard Model, e.g.,
in the Minimal Supersymmetric extension (MSSM)~\cite{br04},
in Grand Unified Theories~\cite{mas06},
and see-saw mechanisms~\cite{ellis}.
In most models describing LFV in the 
$\tau$ lepton sector, 
the radiative decays $\tau\rightarrow\mu(e)\gamma$ have
the largest probability; this has motivated 
many experimental searches~\cite{exp:m2,exp:CB,exp:AG,%
exp:CLEOfirst,exp:DELPHI,exp:CLEOsecond,exp:CLEOthird}.
Using 86 \fbi \ of data recorded at the $\Upsilon(4S)$
resonance with the Belle experiment~\cite{Belle} 
at the KEKB asymmetric-energy $e^+e^-$ collider~\cite{KEKB}
we have previously obtained the upper limits
${\cal B}(\tau^{-}\to\mu^{-}\gamma)<3.1\times10^{-7}$~\cite{TMG} and 
${\cal B}(\tau^{-}\to e^{-}\gamma)<3.9\times10^{-7}$~\cite{TEG}
 at the 90\% confidence
level (CL).
The BaBar collaboration 
subsequently obtained the upper limits
${\cal B}(\tau^{-}\to\mu^{-}\gamma)<6.8\times10^{-8}$~\cite{TMG.babar} and 
${\cal B}(\tau^{-}\to e^{-}\gamma)<1.1\times10^{-7}$~\cite{TEG.babar}
 with 232.2 \fbi \ of data.
Here, we report on an updated analysis with 535 \fbi \ of data,
which corresponds to $4.77\times10^8$ produced 
$\tau^+\tau^-$ pairs~\cite{KKMC}.

The Belle detector is a large-solid-angle magnetic
spectrometer that
consists of a silicon vertex detector (SVD),
a 50-layer central drift chamber (CDC), an array of
aerogel threshold Cherenkov counters (ACC), 
a barrel-like arrangement of time-of-flight
scintillation counters (TOF), and an electromagnetic calorimeter
comprised of CsI(Tl) crystals (ECL) located inside 
a superconducting solenoid coil that provides a 1.5~T
magnetic field.  An iron flux-return located outside
the coil is instrumented to detect $K_L^0$ mesons and to identify
muons (KLM).  The detector
is described in detail elsewhere~\cite{Belle}.

The basic analysis procedure is similar to our previous one \cite{TMG,TEG}.
The selection criteria for 
$\tau^{-}\rightarrow\mu^{-}\gamma/e^{-}\gamma$
are determined and optimized by examining
Monte Carlo (MC) simulated signal and background (BG) events,
including generic $\tau^+ \tau^-$,
$q\bar{q}$ $(q=u,d,s,c,b)$,
 Bhabha, $\mu^{+}\mu^{-}$, and two-photon events.
The BG $\tau^{+}\tau^{-}$ events are generated by the
KKMC/TAUOLA~\cite{KKMC} and
the response of the Belle detector is simulated by the GEANT3~\cite{GEANT3}
based program.

Photon candidates are selected from ECL 
clusters that are 
consistent with an electromagnetic shower shape and 
not associated with charged tracks. 
Muon candidates are identified by using a muon likelihood ratio,
${\cal L}_{\mu}$~\cite{muonID}, which is based on the difference between 
the range of the track
calculated from the particle momentum and that measured
by the KLM. 
This ratio includes the value of 
$\chi^2$ formed from the KLM hit
locations with respect to the extrapolated track.
The muon identification efficiency
for these selection criteria (${\cal L}_{\mu}>0.95$)
is 90\%,
with a pion fake probability of 0.8\%.
Identification of electrons uses
an electron likelihood ratio, 
${\cal L}_e$, based on $dE/dx$ information
from the CDC, the ratio of the energy
deposited in the ECL to the
momentum measured by the CDC and SVD, 
the shower shape in the ECL, 
hit information from the ACC
and matching between the positions of the charged track
and the ECL cluster~\cite{electronID}.
The electron identification efficiency
for these selection criteria (${\cal L}_e>0.9$)
is 95\%, with a pion fake probability of 0.07\%.

\section{\boldmath $\tau\rightarrow\mu\gamma$}

\subsection{Event Selection}

We select events that include exactly two oppositely charged tracks and 
at least one photon, consistent with $\tau^+\tau^-$ decays: 
one $\tau^{\pm}$ (signal side) decays to $\mu^{\pm} \gamma$ and the other 
(tag side) decays to a charged particle that is not a muon 
(denoted hereafter as $\notM$), neutrino(s) and any number of photons.
In this analysis, before particle identification,
we assign the pion mass to all charged tracks 
to calculate their energies
and momenta in the center-of-mass (CM) frame.

\begin{figure}[h]
\begin{center}
\begin{tabular}{cc}
\includegraphics[width=0.45\textwidth]{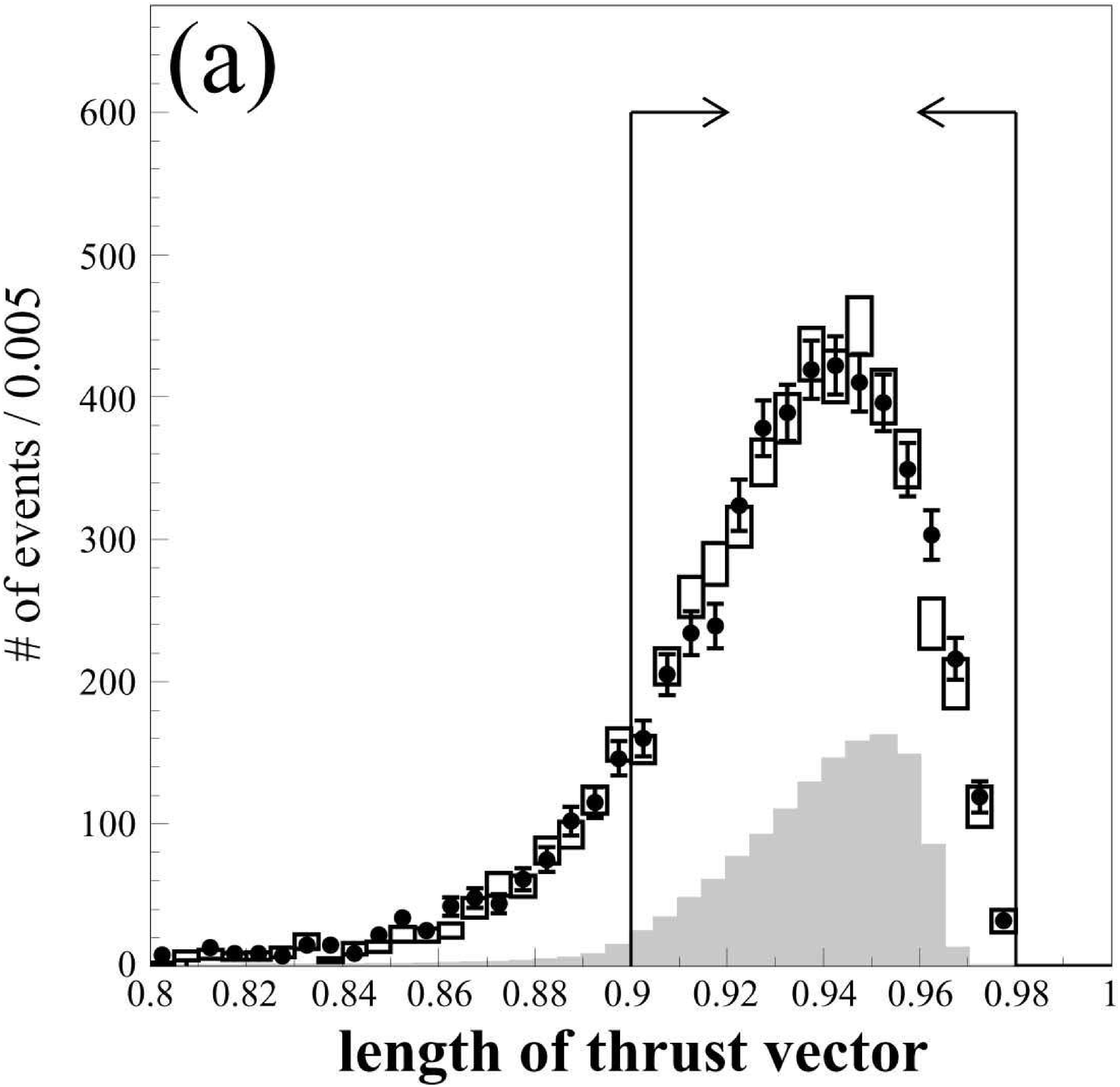} &
\includegraphics[width=0.45\textwidth]{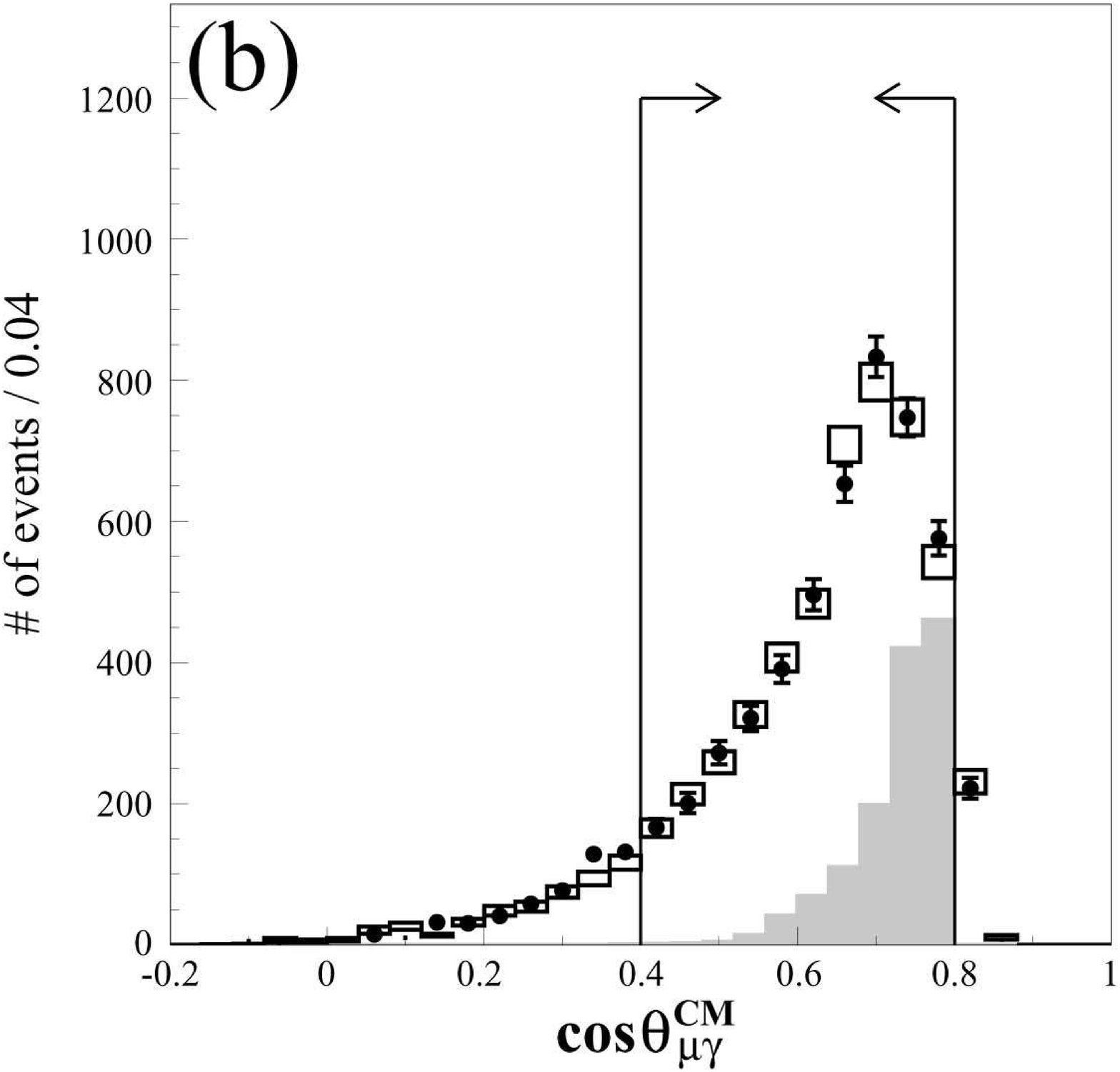}\\
\includegraphics[width=0.45\textwidth]{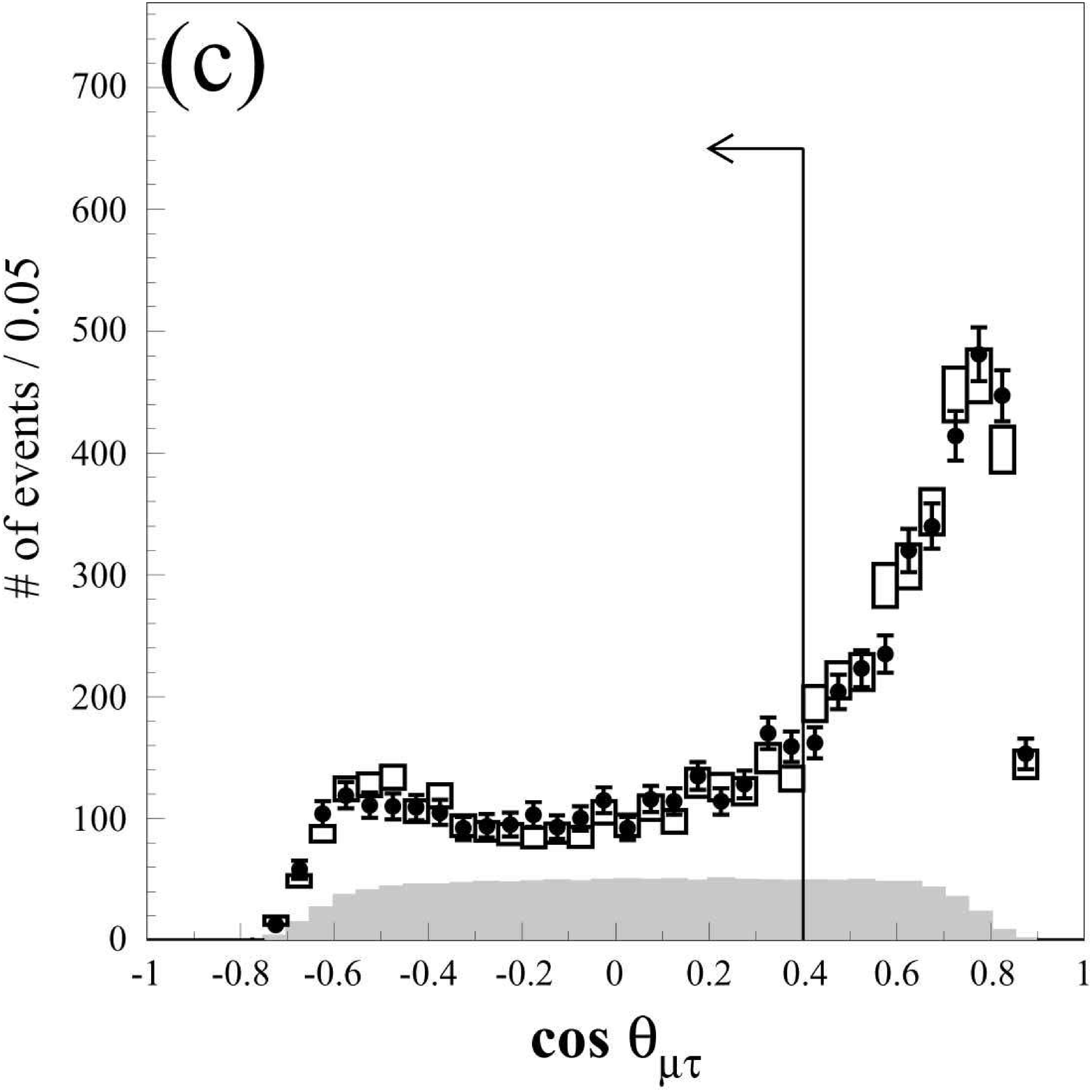}  &
\includegraphics[width=0.45\textwidth]{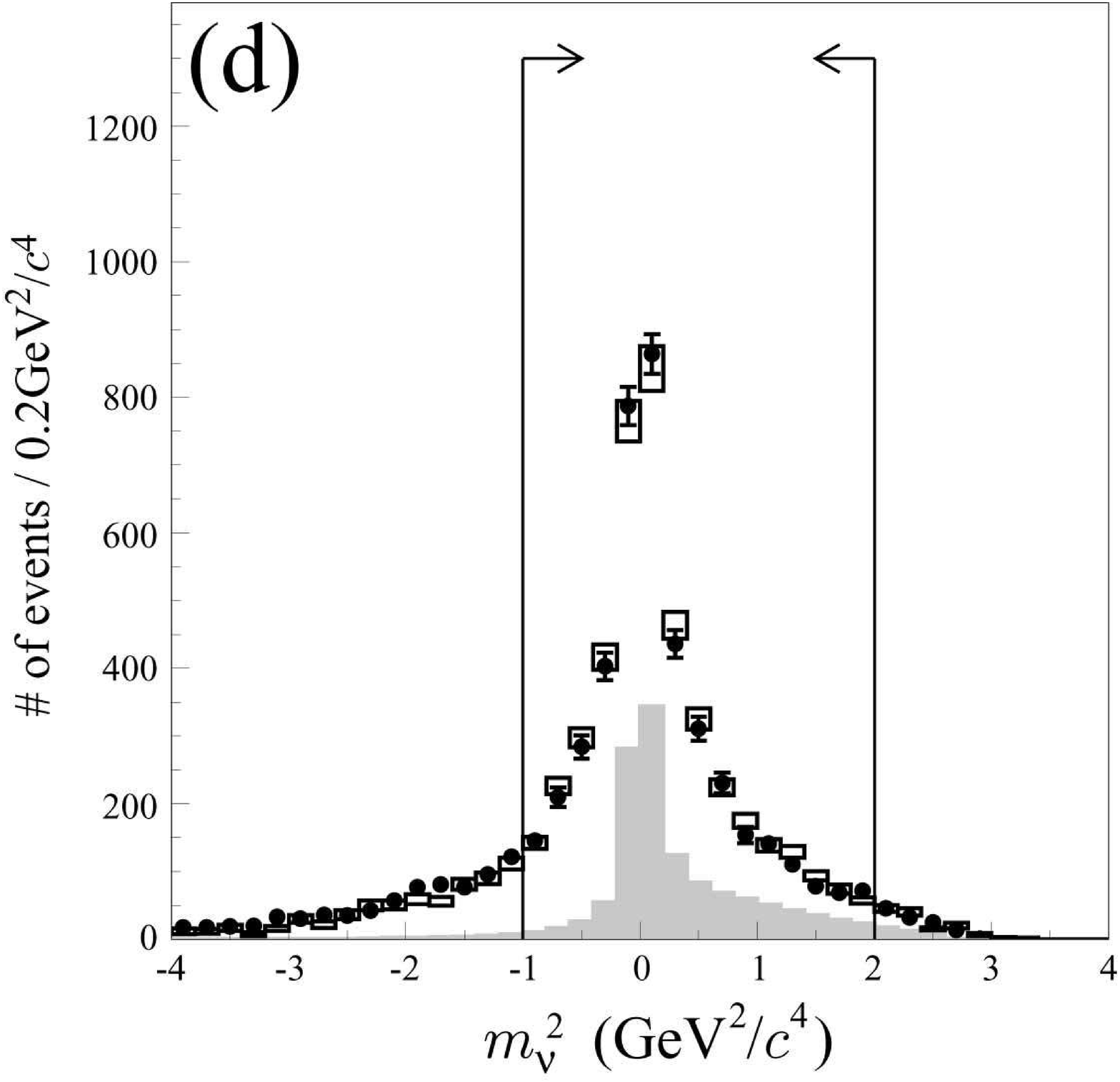} \\
\end{tabular}
\end{center}
\caption{
(a) Magnitude of the thrust vector,
(b) $\cos\theta^{\rm CM}_{\mu\gamma}$,
(c) $\cos\theta_{\mu\tau}$ and
(d) $m^2_{\nu}$ distributions
for $\TMG$. Dots are data, open boxes show the BG MC distribution
and shaded histograms are the signal MC.
Arrows indicate the selected region. 
Here, the branching ratio of
$\tau\rightarrow\mu\gamma$ is assumed to be $1.0\times10^{-5}$ for
the signal MC histograms.
The requirements on particle identification
for the signal and tag sides and $1.5$ GeV/$c^2$ $<\minv<$ 2.0  GeV/$c^2$,
 $-0.5$ GeV $<\dE<$ 0.5 GeV are imposed here.
}
\label{cutmg}
\end{figure}

Each track must have a momentum $p^{\rm CM} <$ 4.5 GeV/$c$ 
in the CM frame 
to reduce contamination
from Bhabha and $\mu^+\mu^-$ events
and a transverse component to the beam axis of
$p_t >$ 0.1 GeV/$c$ within 
the detector fiducial region, 
$ -0.866 < \cos\theta < 0.956$, to suppress two-photon background.
Here $\theta$ is the polar angle with respect
 to the $z$ axis, which is anti-parallel to the $e^+$ beam.
(Hereafter, all the variables defined in the CM frame have 
superscripts ``CM''.)
Each photon is required to have an energy $E_{\gamma}> 0.1$ GeV within 
the fiducial region. 
The total energy in the CM frame must be $E_{\rm total}^{\rm CM} < 10.5$ GeV 
to further suppress Bhabha and $\mu^{+}\mu^{-}$ events.
The magnitude of the thrust vector,
constructed from all selected charged tracks and photons mentioned above,
is required to be in the range from 0.90 to 0.98 
in order to suppress $\mu^{+}\mu^{-}$ and $q\bar{q}$ background
(Fig.~\ref{cutmg}(a)).

For muon identification,
we require a likelihood ratio of 
${\cal L}_{\mu} >$ 0.95 and $p >$ 1.0~GeV/$c$. 
On the tag side, a track with ${\cal L}_{\mu} <$ 0.8 is defined as $\notmu$.
The photon that forms a $(\MG)$ candidate is required to have 
$E_{\gamma} >$ 0.5 GeV and $-0.602 < \cos\theta_{\gamma} < 0.829$ to 
remove any spurious combinations of $\gamma$'s. 

The opening angle between the $\mu$ and $\gamma$ 
of the $(\mu\gamma)$ candidate, 
0.4 $< \cos\theta^{\rm CM}_{\MG} <$ 0.8, 
is useful in rejecting
$\tau^+\tau^-$ background that contains $\pi^0$'s from $\tau$ decays
(Fig.~\ref{cutmg}(b)).
The sum of the energies 
of the two charged tracks and 
the photon of the $(\MG)$ candidate, 
$E^{\rm CM}_{\rm sum}$, should be less than 9.0 GeV to reject 
$\mu^+\mu^-$ events. 
The opening angle between the two tracks should be greater than 90$^{\circ}$ 
in the CM frame, and the opening angle between the $\mu$ 
and the boost direction of its mother $\tau$ from the CM frame
is required to satisfy
$\cos\theta_{\mu\tau}<0.4$ in the $\tau$ rest frame,
to remove combinations of $\mu$'s and $\gamma$'s from BG
(Fig.~\ref{cutmg}(c)).

\begin{figure}[h]
\begin{center}
\includegraphics[width=0.45\textwidth]{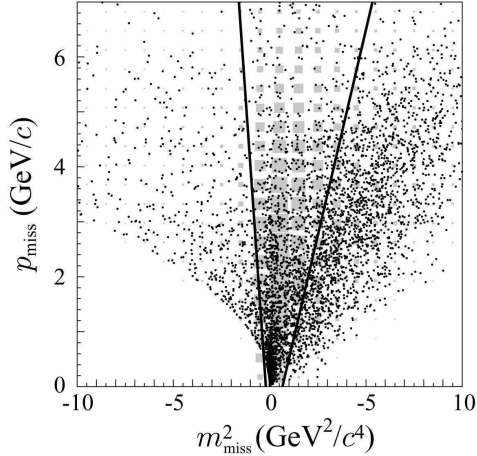} 
\end{center}
\caption{
Distributions in $m^2_{\rm miss}$ -- $p_{\rm miss}$ plane
for data (dot) and signal MC (shaded box) events.
Events are chosen between the two lines in the figure. 
Also, $p_{\rm miss}>0.4$ is required.
The two lines correspond to the equations 
$p_{\rm miss} = -5 \times m^2_{\rm miss}-1$ 
and $p_{\rm miss} = 1.5 \times m^2_{\rm miss}-1$.
The applied requirements are the same as those in Fig. 1.}
\label{vcut}
\end{figure}

The following constraints on the momentum and the polar angle of the 
missing particle are imposed: 
$p_{\rm miss} >$ 0.4 GeV/$c$ and $-0.866 < \cos{\theta}_{\rm miss} <
0.956$. Here, $p_{\rm miss}$ is calculated by subtracting 
the sum of the momenta
of all charged tracks and photons from the beam momenta.
To remove the $\tau^+\tau^-$ background events,
a requirement on 
the opening angle between the tag-side track and the missing particle 
is applied,
0.4 $< \cos\theta^{\rm CM}_{{\rm miss}-\notM} <$ 0.98.
We calculate the missing mass squared on the tag side, $m_{\nu}^2=
(E_{\mu\gamma}^{\rm CM}-E_{\rm tag}^{\rm CM})^2
-(p_{\rm miss}^{\rm CM})^2$, 
where $E_{\mu\gamma}^{\rm CM}$ ($E_{\rm tag}^{\rm CM}$)
is the sum of the energy of the signal side $\mu$ and $\gamma$
in the CM frame,
and then require $-1.0$ GeV${}^2$/$c^4$
$<m_{\nu}^2<2.0$ GeV${}^2$/$c^4$, as shown in Fig.~\ref{cutmg}(d). 
Finally, the condition, 
$p_{\rm miss} > -5 \times m^2_{\rm miss}-1$ and 
$p_{\rm miss} > 1.5 \times m^2_{\rm miss}-1$,
is imposed.
The missing mass is given by
$m^2_{\rm miss}=E^2_{\rm miss} - p^2_{\rm miss}$ (in GeV/$c^2$), 
where $E_{\rm miss}$ is the sum of the beam energies minus the
visible energy.
This relation is illustrated by Fig.~\ref{vcut}. 

\subsection{Background contribution}

After the selection requirements described in the previous subsection, 
the dominant BG source is $\tau^+\tau^-$ events
with the decay $\tau^\pm \to \mu^\pm \nu_{\mu} \nu_{\tau}$ 
or $\tau^\pm \to \pi^\pm \nu_{\tau}$ 
where the $\pi$ is misidentified
as a $\mu$ and is then combined with
 a photon from initial state radiation or beam BG.
Another source is the radiative $e^+e^-\rightarrow\mu^+\mu^-$ process.

Two variables are used to identify the signal: 
$\minv$, 
the invariant mass of ($\mu\gamma$), and 
$\Delta E = E_{\mu\gamma}^{\rm CM} 
- E^{\rm CM}_{\rm beam}$, the energy difference between the $(\mu\gamma)$ 
energy and the beam energy in the CM frame,
where the signal should have $\minv\sim$ $m_{\tau}$ and $\dE\sim0$. 
The resolutions in $\minv$ and $\dE$ are estimated by fitting 
 asymmetric Gaussians to 
the signal MC distributions giving
$\sigma^{\rm high/low}_{\minv}=14.49\pm0.10/24.24\pm0.13$ 
\mevpcs \ and $\sigma^{\rm high/low}_{\dE}=35.29\pm0.49/81.41\pm0.94$
\mev,
where $\sigma^{\rm high/low}$ means the standard deviation on the higher/lower
side of the peak.

To compare the data and MC simulation, we examine a $5\sigma$ region with
$1.65$ \gevpcs \ $<\minv<1.85$ \gevpcs \ and $-0.41$ \gev \
$<\dE<0.17$ \gev, as shown in 
Fig.~\ref{2d_mg}(a). 
A `blind analysis' method is used: 
a $3\sigma$ region 
as indicated by the dashed ellipse in Fig.~\ref{2d_mg}(a)
is `blinded' (not examined) 
until all selection criteria are finalized.
The detection efficiency for this region is determined
from MC simulation to be $6.05\%$.

After the selections,
we find 71 events remaining in data and 
73.4$\pm 6.7$ events in MC in the $5\sigma$ region
 outside the blinded ellipse. 
According to MC,
the remaining events are
 dominated by the initial state 
radiation process $\tau^+ \tau^- \gamma$
 --  $58.8\pm 4.3$ $(70.3\pm 4.7)$ events 
and also include
$13.1\pm 4.9$ $(15.0\pm 5.3)$  $\mu^{+}\mu^{-}\gamma$ events
with incorrect $\mu$ identification,
and $1.6\pm 1.6$ $(3.2\pm 2.2)$  two-photon events,
where the numbers 
in parentheses are the BGs that remain 
in the entire $5\sigma$ region.

This background composition was understood 
in the previous analysis; 
the $\tau^{+}\tau^{-}\gamma$ process yields contributions in the 
$\Delta E<0$ region, while 
$\mu^{+}\mu^{-}\gamma$ events mostly have $\Delta E>0$. 
This BG distribution is well represented by a combination of Landau and 
Gaussian functions, as found in Ref.~\cite{TEG}.
To obtain the final BG distribution,
we perform a binned maximum likelihood fit to the candidates
in the $5\sigma$ region outside the blinded ellipse.
The final event distribution in the data is very similar
to that obtained from the MC: $(79.0\pm 9.2)$\% is $\tau^+\tau^-\gamma$,
$(15.8\pm 8.5)$\% is $\mu^+\mu^-\gamma$, and $(5.2\pm 4.1)$\% is from
$e^+e^-\gamma\rightarrow e^+e^-\mu^+\mu^-$.

\subsection{Signal extraction}

\begin{figure}[h]
\center{ \includegraphics[width=0.4\textwidth]{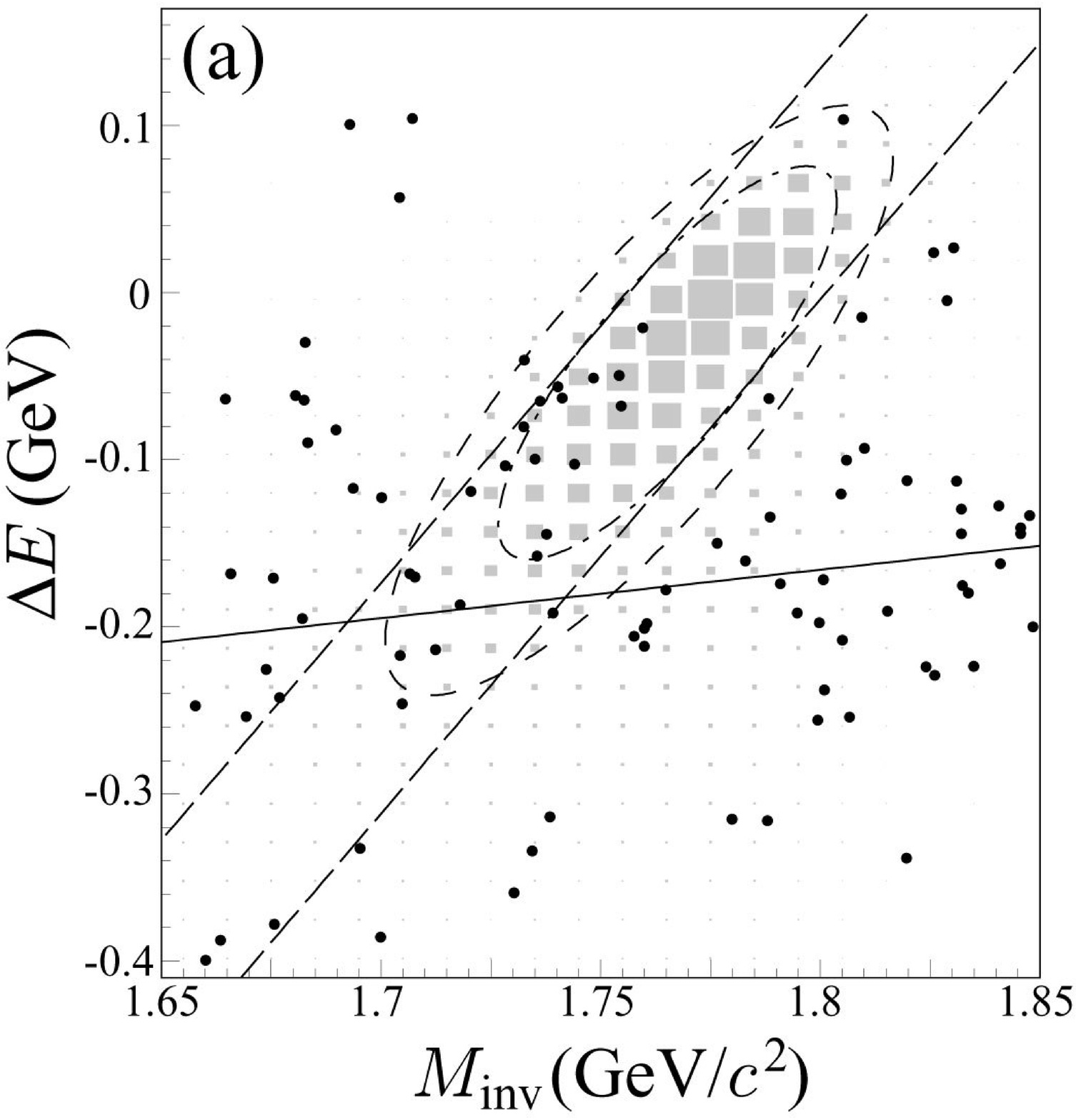}~~
         \includegraphics[width=0.4\textwidth]{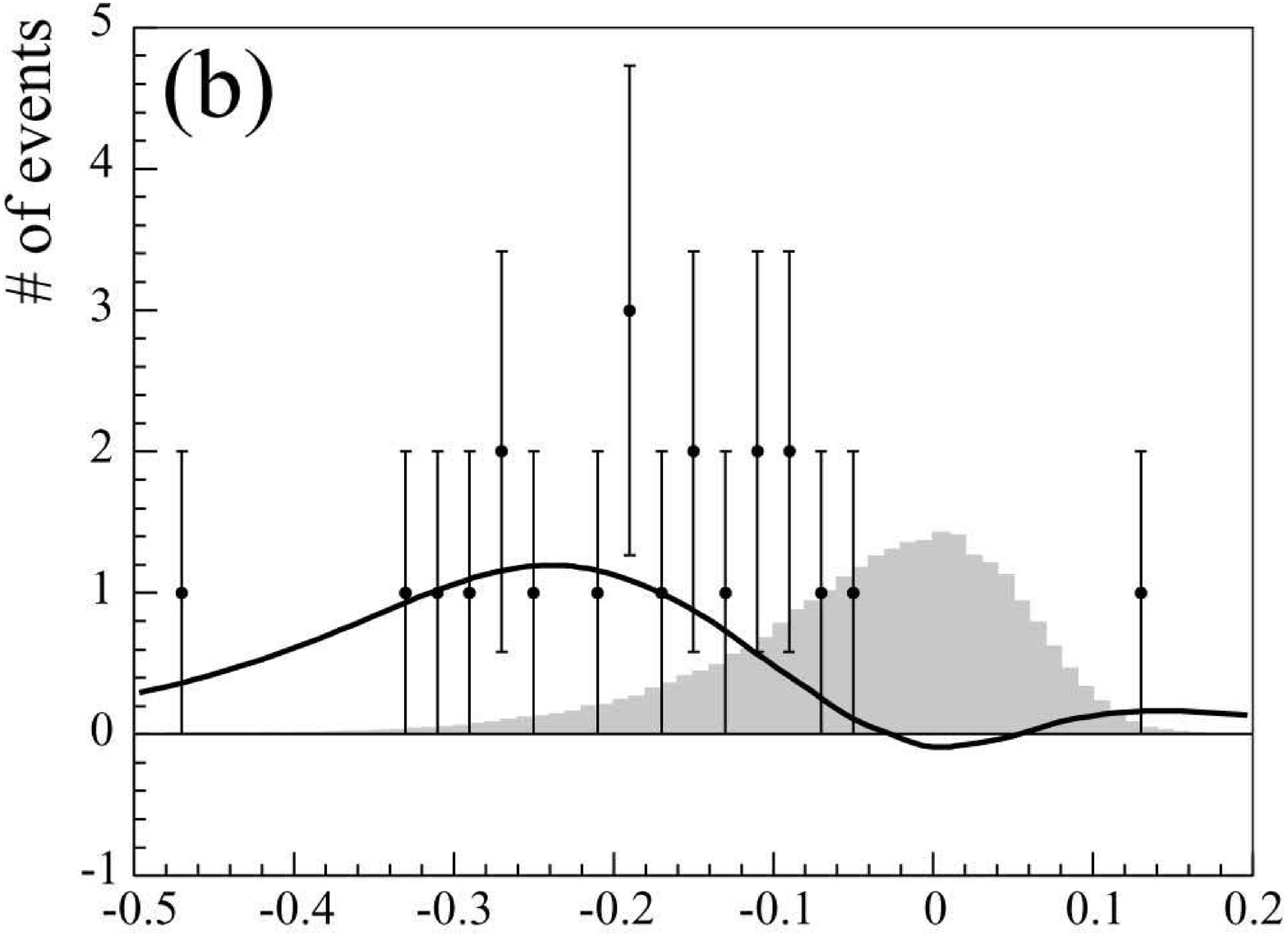} }
\caption{(a) $\minv$ -- $\dE$ distribution in the search for
$\TMG$ in the $5\sigma$ region.
Dots are the data and shaded boxes indicate the signal MC.
The dashed ellipse shows the $3\sigma$ blinded region
and the dot-dashed ellipse is the $2\sigma$ signal region.
The dashed lines indicate the $2\sigma$ band of the shorter 
ellipse axis, projected onto the longer ellipse axis.
The solid line indicates the dense BG region.
(b) Data distribution within the $2\sigma$ band.
Here, $\cal{E}\equiv\dE-\dE^{(0)}$,
$\cal{M}\equiv3.0\times{c^2}\times(\minv-\minv^{(0)})$ and
 $\alpha=46^{\circ}$. We obtain the most probable values 
$\dE^{(0)}$ and $\minv^{(0)}$,
$\minv^{(0)}=1.776$ GeV/$c^2$,
$\dE^{(0)}=-5$ MeV,
by fitting the signal MC distribution to an asymmetric Gaussian.
Points with error bars 
are the data and the shaded histogram is the signal MC
assuming a branching ratio of $5\times10^{-7}$.
The solid curve shows the best fit.
}
\label{2d_mg}
\end{figure}

After unblinding, we find 23 and 94 data events in the blinded and 
$5\sigma$ regions, respectively, while $15.0\pm3.1$
and $88.4\pm7.4$ events are expected from the MC. 
Figure~\ref{2d_mg}(a) shows the final event distributions 
(data and signal MC)
in the $\minv$--$\dE$ plane. 

In order to extract the number of signal events, 
we employ an unbinned extended maximum 
likelihood (UEML) fit  with the following likelihood function: 
\begin{equation}
  {\cal L}=\frac{e^{-(s+b)}}{N!}\prod_{i=1}^{N}
   \left(s S_i+b B_i \right).
\end{equation}
Here, $N$ is the number of observed events;
 $s$ and $b$ are the numbers of signal 
and BG events to be extracted, respectively; $S_i$ and $B_i$ are 
the signal and BG probability density functions (PDF),
where $i$ indicates the $i$-th event; 
the shape of the signal PDF, $S_i$, 
is obtained by smoothing the signal MC distribution, and 
$B_i$ is the PDF for the background mentioned above, whose distribution is
concentrated around $\Delta E\simeq -0.2$ GeV, as indicated by 
the solid line in Fig.~\ref{2d_mg}(a). 
To enhance the signal detection sensitivity 
and to avoid this dense BG region, 
we use a $ 2\sigma$ ellipse as the signal region for the UEML fit. 
The result of the fit is $s=-3.9^{+3.6}_{-3.2}$, 
$b=13.9^{+6.0}_{-4.8}$ with $N=10$.

Figure~\ref{2d_mg}(b) shows the event distribution within 
the $ 2\sigma$ band 
of the shorter ellipse axis, projected onto the longer ellipse axis, and 
the best fit curve. 
No events are found near the peak of the signal distribution. 
The negative $s$ value is consistent with no signal.

We examine the probability to obtain this result and evaluate the 90\% 
CL upper limit using a toy MC simulation.
The toy MC generates signal and BG events according to their PDFs 
fixing the expected number of BG events $(\tilde{b})$ at $\tilde{b}=b$, 
while varying the number of signal events ($\tilde{s}$). 
For every assumed $\tilde{s}$,  10,000 samples are generated following 
Poisson statistics with means $\tilde{s}$ and $\tilde{b}$ for the signal 
and BG, respectively; 
the signal yield $(s^{\rm MC})$ is evaluated by the UEML fit. 
To obtain the upper limit at the 90\% CL ($\tilde{s}_{90}$) we take 
an $\tilde{s}$ value that gives a 10\% probability
for $s^{\rm MC}$ to be smaller 
than $s$.  
The probability to obtain $s\leq -3.9$ is 25\% in the case of a null 
true signal.
In other words, due to BG fluctuations
a negative $s$ value is possible with a large probability, 
although the physical signal rate is positive~\cite{OLDadd}.

The toy MC provides an upper limit on the signal at the 90\% CL as 
$\tilde{s}_{90}=2.0$ events from the UEML fit. 
We then obtain the upper limit on the branching ratio
${\cal B}_{90}(\tau\rightarrow\mu\gamma)$ 
at the 90\% CL as
\begin{equation}
 {\cal B}_{90}(\tau^-\rightarrow\mu^-\gamma)\equiv
\frac{\tilde{s}_{90}}{2\epsilon N_{\tau\tau}} = 4.1\times 10^{-8},
\end{equation}
where the number of $\tau$ pairs produced is $N_{\tau\tau} =
 (4.77\pm0.07) \times 10^8$,
and the detection efficiency for the $ 2\sigma$ ellipse region is 
$\epsilon= 5.07\%$.

The systematic uncertainties for the BG PDF shape increase
$\tilde{s}_{90}$ to 2.2~\cite{CLEO.mgeg}.
The other systematic uncertainties arise from
the track reconstruction efficiency (2.0\%), 
the photon reconstruction efficiency (2.0\%), 
the selection criteria (2.2\%), 
the luminosity (1.4\%), 
the trigger efficiency (0.9\%), 
and the MC statistics (0.3\%). 
All contributions are added in quadrature to obtain
the total uncertainty of 4.0\%. 
This uncertainty increases the upper limit on the branching ratio by
0.2\%~\cite{CLEO.mgeg}.
Since the angular distribution for $\tau\to\mu\gamma$ depends on 
the LFV interaction 
structure, we evaluate its effect on the result by assuming the maximum 
possible variation, $V\pm A$ interactions,
rather than the uniform distribution that is the default 
in the MC analysis. 
No appreciable effect is found for the upper limit.

Finally, the following upper limit on the branching ratio is obtained: 
\begin{equation}
 {\cal B}(\tau^-\rightarrow\mu^-\gamma)<4.5\times10^{-8} 
\hspace*{1 cm} {\rm at~the~90\% ~CL.}
\end{equation}
\vspace*{2 mm}

\section{\boldmath $\tau\rightarrow{e}\gamma$}

For $\tau\rightarrow e\gamma$ 
we use a procedure similar to that for 
$\tau\rightarrow\mu\gamma$. 

\subsection{Event Selection}

We examine a $\tau^+\tau^-$ sample, 
in which one $\tau$ decays to 
an electron and a photon,
and the other $\tau$ decays to a charged particle, 
but not an electron ($\notE$), neutrino(s) and any number of photons.
Because the selection criteria 
are quite similar to those for $\tau\to\mu\gamma$, 
below we describe only the differences.

\begin{figure}[h]
\begin{center}
\begin{tabular}{cc}
\includegraphics[width=0.45\textwidth]{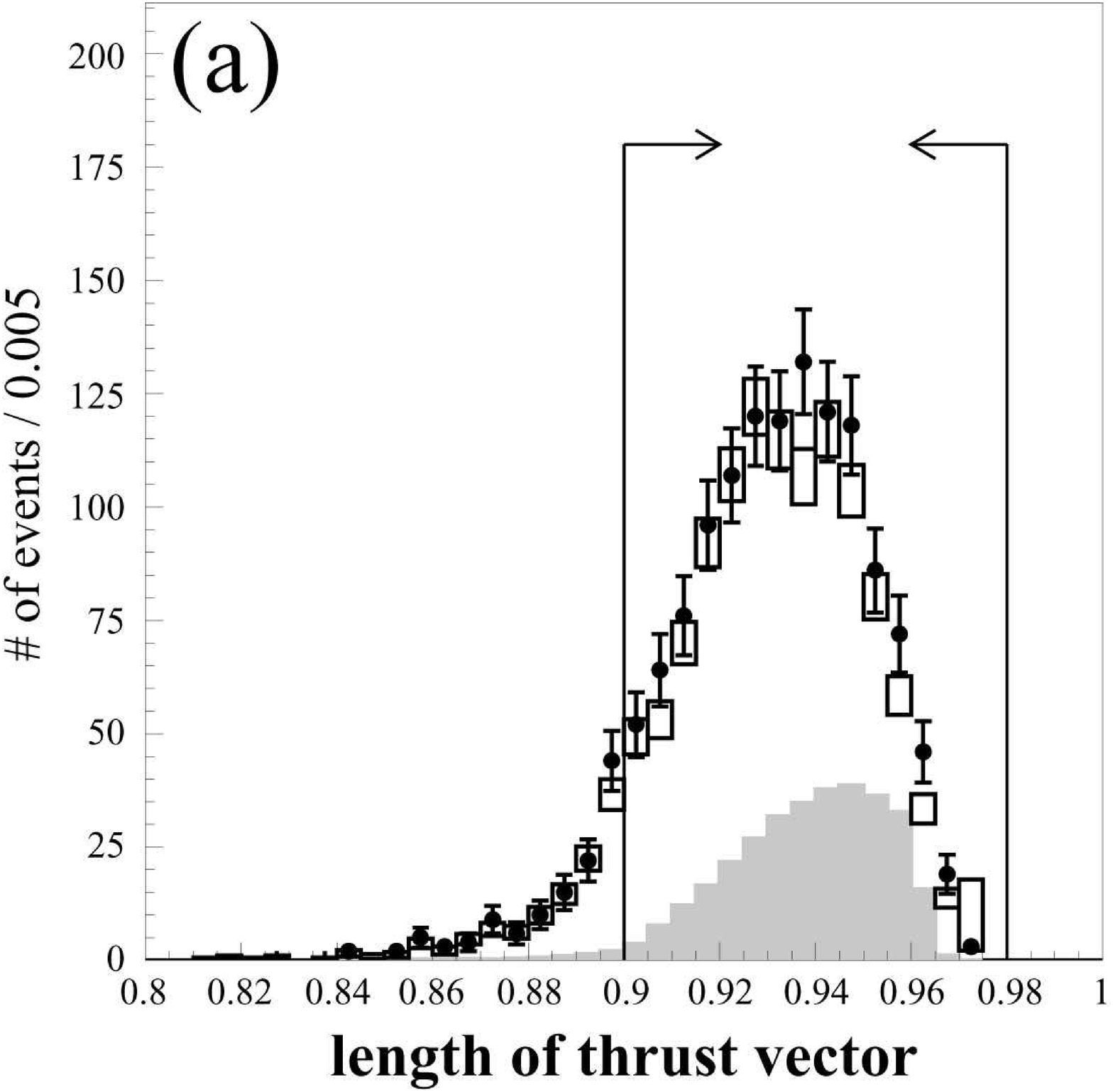} &
\includegraphics[width=0.45\textwidth]{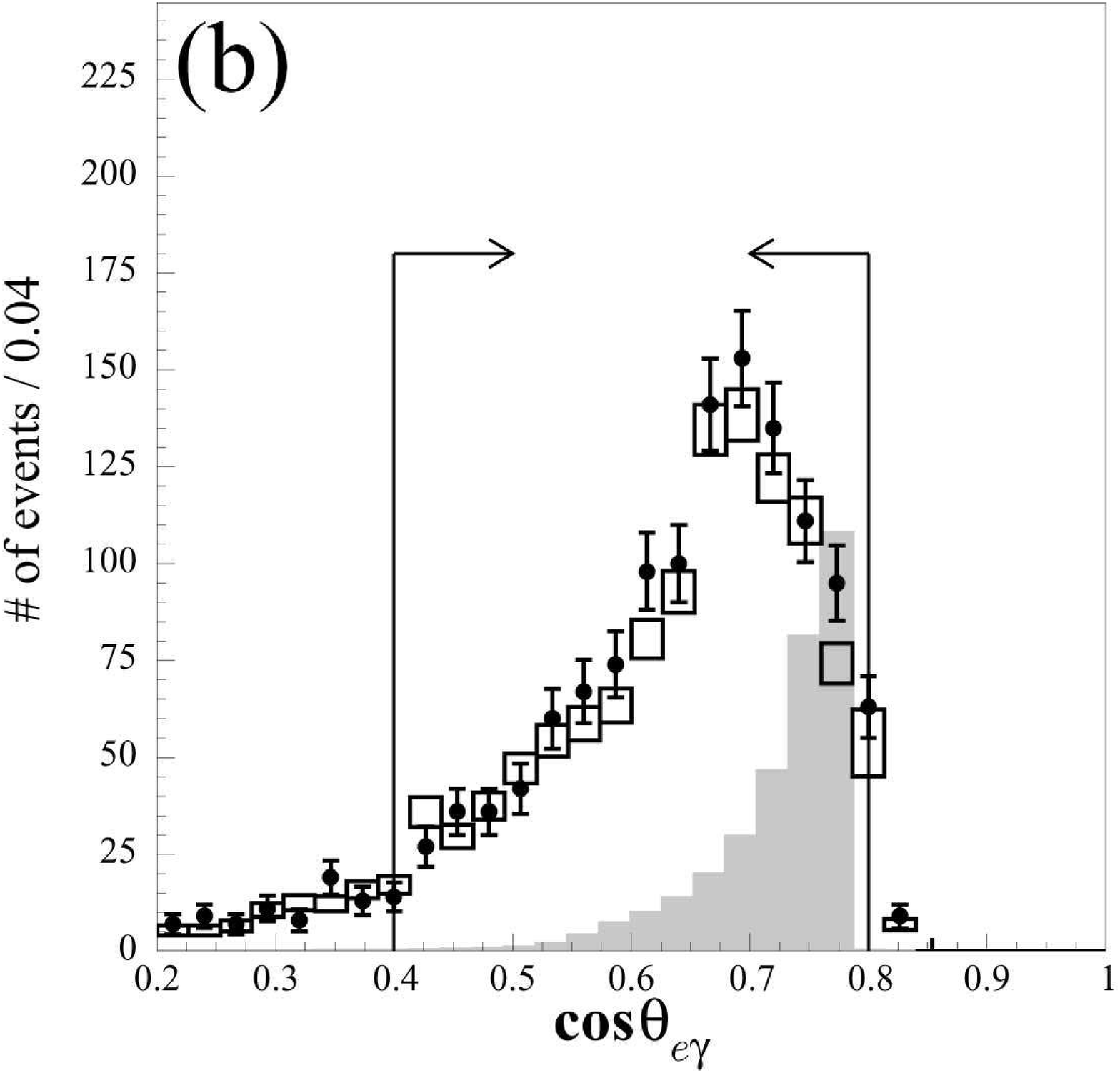}\\
\includegraphics[width=0.45\textwidth]{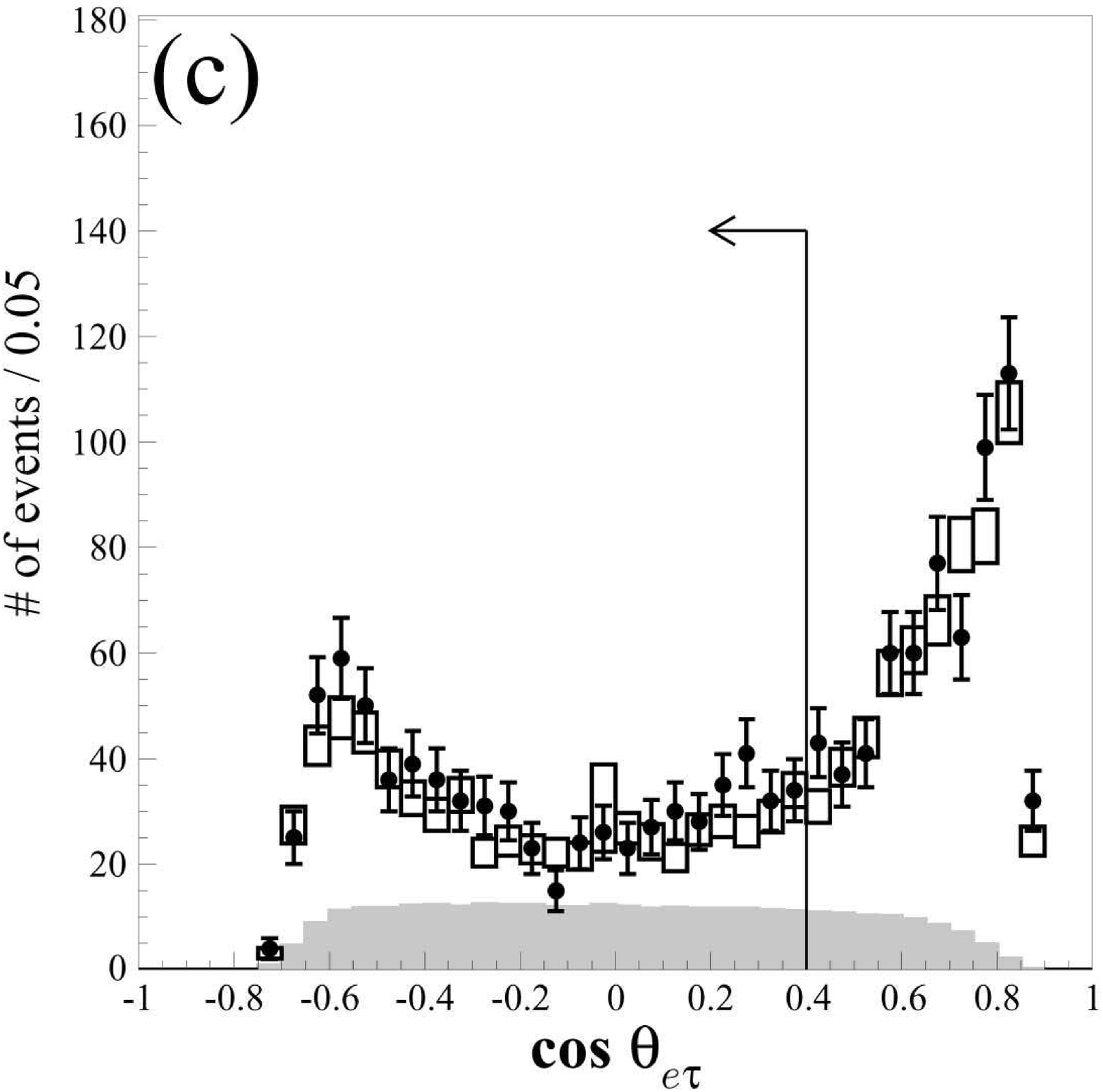}  &
\includegraphics[width=0.45\textwidth]{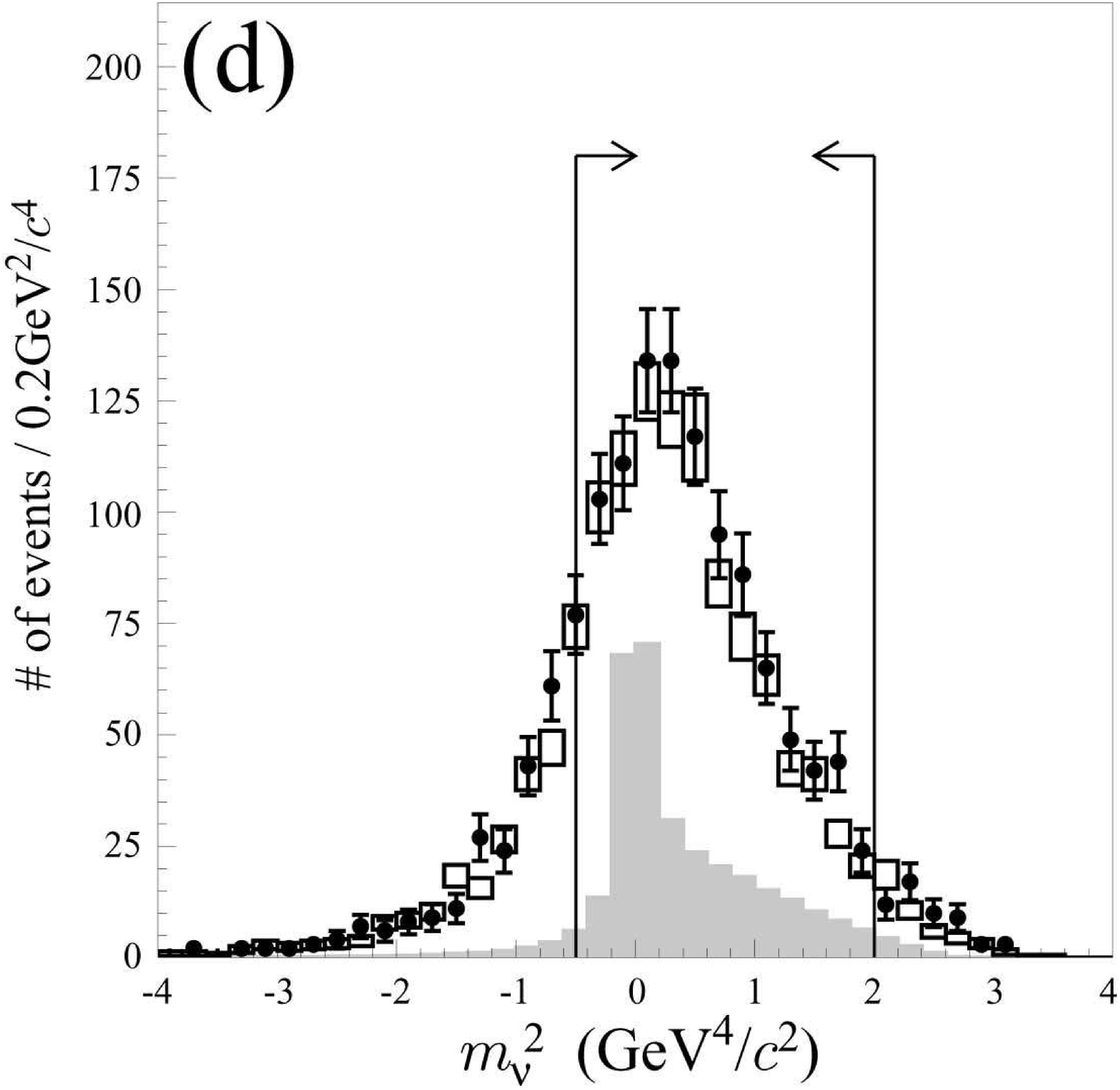} \\
\end{tabular}
\end{center}
\caption{
(a) Magnitude of the thrust vector,
(b) $\cos\theta^{\rm CM}_{e\gamma}$,
(c) $\cos\theta_{e\tau}$ and
(d) $m^2_{\nu}$ distributions
for $\TEG$. Dots are data, 
open boxes show the BG Monte Carlo(MC) distribution
and shaded histograms are the signal MC.
Arrows indicate the selected region. 
Here, the branching ratio 
of $\tau\rightarrow{e}\gamma$ is assumed to be $5.0\times10^{-6}$ 
for the signal MC histograms.
The requirements on particle identification
for the signal and tag sides and $1.5$ GeV/$c^2$ $<\minv<$ 2.0  GeV/$c^2$,
 $-0.5$ GeV $<\dE<$ 0.5 GeV
were applied 
for these figures.
}
\label{cuteg}
\end{figure}

An obvious difference is the replacement of
the $\mu$ by an $e$ on the signal side, 
and using an $e$ veto ($\notE$) rather than 
a $\mu$ veto ($\notM$) on the tag side.
The electron on the signal side $(\EG)$ is required to have 
${\cal L}_{\rm e} >$ 0.90 and a momentum $p >$ 1.0 GeV/$c$, while the $\notE$ 
on the tag side should have ${\cal L}_{\rm e} <$ 0.1.
Minor differences in the kinematical selection include requirements
on the missing 
mass squared 
on the tag side and the opening angle between the tag-side track and the 
missing particle on the tag side: $-0.5$ GeV${}^2$/$c^4$
$< m_{\nu}^2 < 2.0$ GeV${}^2$/$c^4$,
and 0.4 $< \cos\theta^{\rm CM}_{{\rm miss}-\notE} <$ 0.99. 
The other requirements are the same as those for $\tau\to\mu\gamma$,
as shown in Fig.~\ref{cuteg}.

The $M_{\rm inv}$ and $\Delta E$ resolutions are
$\sigma^{\rm high/low}_{\minv}=14.76\pm0.18/25.38\pm0.38$ \mevpcs \ and 
$\sigma^{\rm high/low}_{\dE}=35.66\pm0.62/89.98\pm1.72$ \mev.
The $5\sigma$ region with
$1.65$ \gevpcs \ 
$<\minv<1.85$ \gevpcs \ and $-0.45$ \gev \ $<\dE<0.18$ \gev \ 
is used for the signal evaluation. 
A $3\sigma$ ellipse is also blinded. 

After the selection requirements
we find 42 and $34.7\pm3.3$ events
outside the blind in the $5\sigma$ region
in data and MC, respectively.
As our MC study shows, 
the BG comes from $\tau^{+}\tau^{-}\gamma$ events.

\subsection{Signal extraction}

\begin{figure}[h]
\center{ \includegraphics[width=0.4\textwidth]{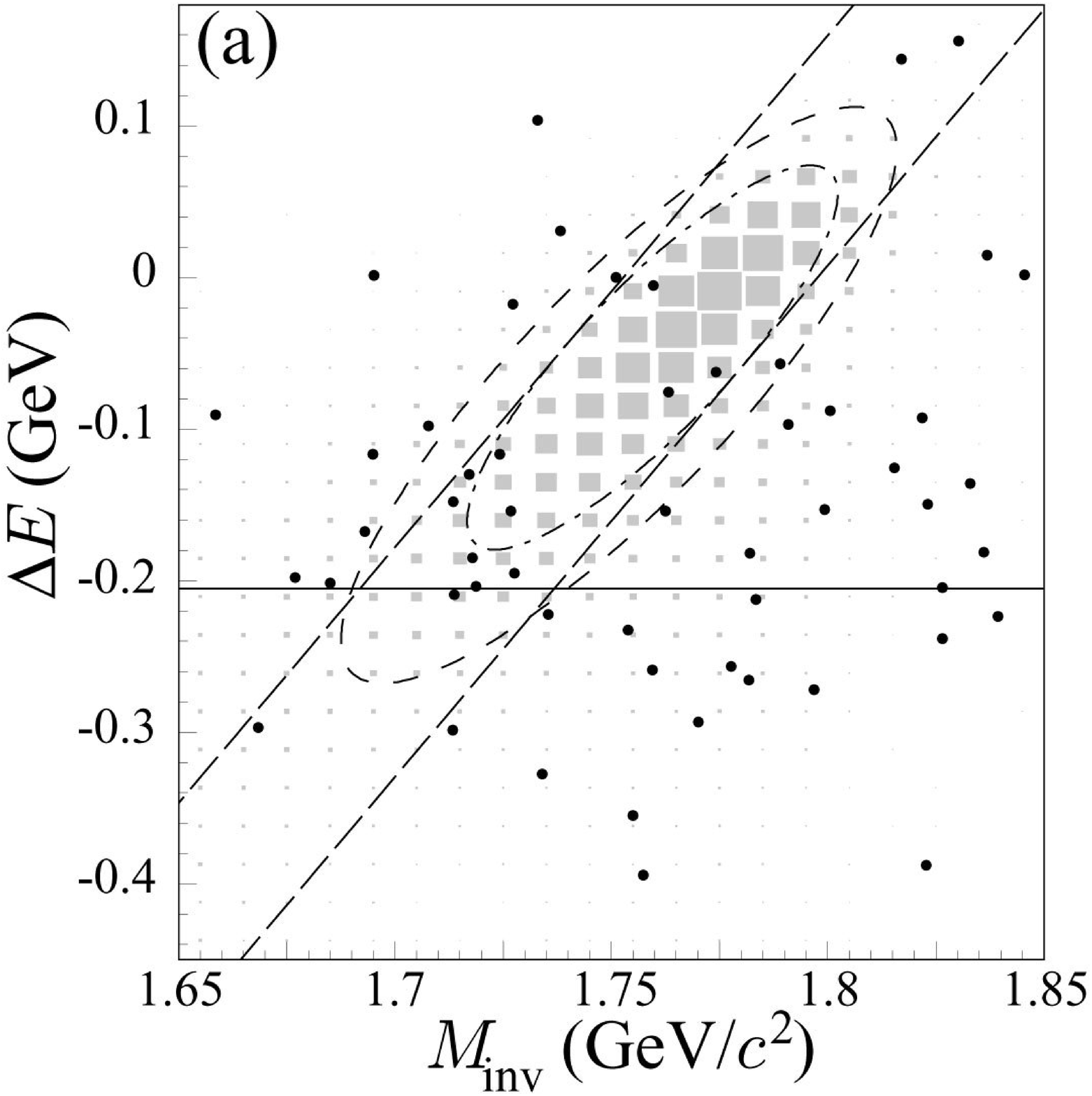}~~
         \includegraphics[width=0.4\textwidth]{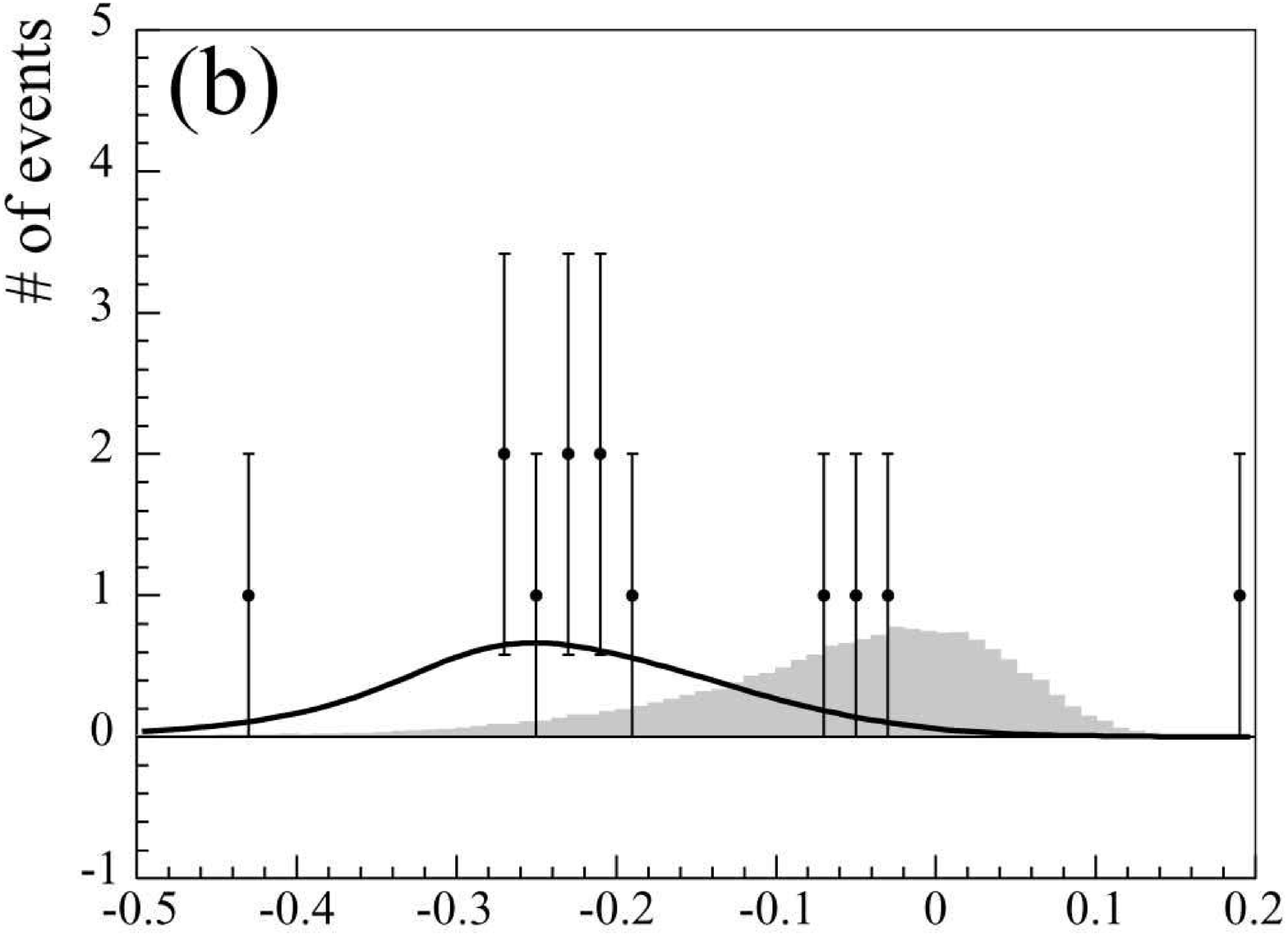} }

\caption{(a) $\minv$ -- $\dE$ distribution in the search for
$\TEG$ in the $5\sigma$ region.
Dots are the data and shaded boxes indicate the signal MC.
The dashed ellipse shows the $3\sigma$ blinded region
and the dot-dashed ellipse is the $2\sigma$ signal region.
The dashed lines indicate the $2\sigma$ band of the shorter ellipse axis,
projected onto the longer ellipse axis.
The solid line indicates the dense BG region.
(b) Data distribution within the $2\sigma$ band.
Here, $\cal{E}\equiv\dE-\dE^{(0)}$,
$\cal{M}\equiv3.0\times{c^2}\times(\minv-\minv^{(0)})$ and
 $\alpha=48^{\circ}$. We obtain the most probable values 
$\dE^{(0)}$ and $\minv^{(0)}$,
$\minv^{(0)}=1.775$ GeV/$c^2$,
$\dE^{(0)}=-2$ MeV,
by a fit of the signal MC distribution to an asymmetric Gaussian.
Points with error bars are the data and 
the shaded histogram is the signal MC
assuming a branching ratio of $5\times10^{-7}$.
The solid curve shows the best fit.
}
\label{2d_eg}
\end{figure}

After opening the blind we find 13 and 55 data events in the blinded
region and 
$5\sigma$ region, respectively, while the MC predicts
$8.1\pm1.6$ and $42.8\pm 3.7$ events, respectively. 
Figure~\ref{2d_eg}(a) shows the event distribution 
in the $\minv$--$\dE$ plane. 

The signal extraction process is the same as that for $\tau\to\mu\gamma$, 
described in the former section. 
The BG is composed of  $(18 \pm 18)\%$ $e^+e^-\gamma$ (radiative
Bhabha), while the remainder is $\tau^+\tau^-\gamma$.
No other background source is found in MC. 
The UEML fit over the $2\sigma$ ellipse region results
 in $s=-0.14^{+2.18}_{-2.45}$, 
$b=5.14^{+3.86}_{-2.81}$ with $N=5$. 
The toy MC gives a probability of 48\% to obtain $s\leq -0.14$ in 
the case of a null signal. 
Figure~\ref{2d_eg}(b) is the same as Fig.~\ref{2d_mg}(b), but for 
the $\tau\to e\gamma$ case. 
The upper limit of $\tilde{s}_{90}=3.3$ is obtained by the toy MC in 
the case of the UEML fit result.
The upper limit on the branching ratio is calculated as 
\begin{equation}
{\cal B}_{90}(\tau^-\rightarrow e^-\gamma)\equiv
\frac{\tilde{s}_{90}}{2\epsilon N_{\tau\tau}} = 11.7\times 10^{-8},
\end{equation}
where the detection efficiency for the $ 2\sigma$ ellipse region is 
$\epsilon= 2.99\%$. 

The systematic uncertainties for the BG PDF shape increase
$\tilde{s}_{90}$ to 3.4.
The other systematic uncertainties are similar to those for 
$\tau\to\mu\gamma$; minor differences are 
in the selection criteria (2.5\%)
and the trigger efficiency (2.0\%). 
The total uncertainty is 4.5\%, and it increases the upper limit 
on the branching ratio by 0.2\%. 
Taking into account this systematic error, we obtain 
the 90\% CL upper limit,
\begin{equation}
{\cal B}(\tau^-\rightarrow{e^-}\gamma) < 12.0\times10^{-8}.
\end{equation}

\section{Summary}

We have updated our searches for the LFV decay modes,
$\TMG$ and $\TEG$, using
535 fb$^{-1}$ of data,
corresponding to about six times higher statistics than 
in the previous publication~\cite{TMG,TEG}.
In this study,
we have introduced four new requirements for $E^{\rm CM}_{\rm total}$,
$m^2_{\nu}$, $\cos\theta_{\mu\tau}$ and the magnitude of the thrust vector
in the $\tau\rightarrow\mu\gamma$ search
and
three new requirements for
$m^2_{\nu}$, $\cos\theta_{e\tau}$ and the magnitude of the thrust vector
in the $\tau\rightarrow{e}\gamma$ search.
As a result of the optimized selection criteria,
we obtain 
$\epsilon=5.07 \ (2.99)$\%
and $N_{\rm obs}=10 \ (5)$ events
in the $2\sigma$ elliptical signal region
for $\tau\to{\mu}\gamma$ ($e\gamma$) in this analysis,
  compared to $\epsilon=12.0 \ (6.39)$\% 
and $N_{\rm obs}=54 \ (20)$ events over a
$5 \sigma$ signal box in our previous study.
The improved selection leads to a value of 
$\epsilon/\sqrt{N_{\rm obs}}$
that is almost the same as that in the previous analysis,
  and therefore the expected upper limit, which is approximately
  proportional to $\sqrt{N_{\rm obs}}/(\epsilon\int L dt)$ in the 
  case of no signal,
  has improved as $1/\int L dt$ rather than in proportion to
  $1/\sqrt{\int L dt}$, as would be the case if the previous
  selection criteria were used. 
(Here $\int L dt$ is the integrated luminosity.)

The resulting upper limits on the branching ratios are 
\begin{eqnarray}
 {\cal B}(\tau^-\rightarrow\mu^-\gamma) &<& 4.5 \times 10^{-8},\\
 {\cal B}(\tau^-\rightarrow{e^-}\gamma) &<& 12.0 \times 10^{-8}
\end{eqnarray}
at the 90\% CL. \\ 

The limit on the branching ratio of the $\tau^- \to \mu^- \gamma$ decay
is the most stringent among all LFV decays of the $\tau$ lepton to
date. 
Our results constrain the parameter space of various theoretical
models that predict LFV decays~\cite{br04,mas06,ellis}.

\section*{Acknowledgments}

We thank the KEKB group for the excellent operation of the
accelerator, the KEK cryogenics group for the efficient
operation of the solenoid, and the KEK computer group and
the National Institute of Informatics for valuable computing
and Super-SINET network support. We acknowledge support from
the Ministry of Education, Culture, Sports, Science, and
Technology of Japan and the Japan Society for the Promotion
of Science; the Australian Research Council and the
Australian Department of Education, Science and Training;
the National Science Foundation of China and the Knowledge 
Innovation Program of the Chinese Academy of Sciences under 
contract No.~10575109 and IHEP-U-503; the Department of Science 
and Technology of India; the BK21 program of the Ministry of Education of
Korea, and the CHEP SRC program and Basic Research program 
(grant No. R01-2005-000-10089-0) of the Korea Science and
Engineering Foundation; the Polish State Committee for
Scientific Research under contract No.~2P03B 01324; the
Ministry of Science and Technology of the Russian
Federation; the Slovenian Research Agency;  
the Swiss National Science Foundation; the National Science Council and
the Ministry of Education of Taiwan; and the U.S.\
Department of Energy.


\begin{thebibliography}{99}
\bibitem{OLDadd}
We restricted $s$ to be positive in the previous analysis. 
Due to BG fluctuations, however, $s$ can be negative as we found in this 
case. Our previous upper limit is then reduced by a factor of $\simeq 3$ 
when negative $s$ is allowed.
\bibitem{CLEO.mgeg}
  A method to incorporate the systematic uncertainties into a branching
  ratio is discussed in Ref. \cite{exp:CLEOthird}.
\end{thebibliography}
\end{document}